# Contour Generation with Realistic Inter-observer Variation

Eliana Vásquez Osorio[†], Jane Shortall, Jennifer Robbins and Marcel van Herk

*Division of Cancer Sciences; Faculty of Biology, Medicine and Health; The University of Manchester*

Contours are used in radiotherapy treatment planning to identify regions to be irradiated with high dose and regions to be spared. Therefore, any contouring uncertainty influences the whole treatment. Even though this is the biggest remaining source of uncertainty when daily IGRT or adaptation is used, it has not been accounted for quantitatively in treatment planning. Using probabilistic planning allows to directly account for contouring uncertainties in plan optimisation. The first step is to create an algorithm that can generate many realistic contours with variation matching actual inter-observer variation.

We propose a methodology to generate random contours, based on measured spatial inter-observer variation, IOV, and a single parameter that controls its geometrical dependency: $\sigma$, the width of the 3D Gaussian used as point spread function (PSF). We used a level set formulation of the median shape, with the level set function defined as the signed distance transform. To create a new contour, we added the median level set and a noise map which was weighted with the IOV map and then convolved with the PSF. Thresholding the level set function reconstructs the newly generated contour.

We used data from 18 patients from the golden atlas, consisting of five prostate delineations on T2-w MRI scans. To evaluate the similarity between the contours, we calculated the maximum distance to agreement to the median shape (maxDTA), and the minimum dose of the contours using an ideal dose distribution. We used the two-sample Kolmogorov-Smirnov test to compare the distributions for maxDTA and minDose between the generated and manually delineated contours.

Only $\sigma = 0.75\ cm$ produced maxDTA and minDose distributions that were not significantly different from the manually delineated structures. Accounting for the PSF is essential to correctly simulate inter-observer variation. The first step to incorporate contour variations directly into treatment optimisation has been taken.

# Contour Generation with Realistic Inter-observer Variation

Eliana Vásquez Osorio[†], Jane Shortall, Jennifer Robbins and Marcel van Herk
*Division of Cancer Sciences; Faculty of Biology, Medicine and Health; The University of Manchester*

**Introduction**
Tumour and organ at risk delineations are essential when planning radiotherapy: their location and shape determine the regions to receive high dose and those to spare. Any uncertainty in their definition influences the whole treatment, potentially leading to underdosing tumour regions or overdosing organs at risk. Even though these uncertainties have been quantified for several sites and image modalities (i.e. inter- and intra-observer variation), they are typically not accounted for quantitatively in treatment planning or margin calculation. This is becoming ever more relevant, as with daily IGRT or adaptation, delineation variation is the biggest remaining source of uncertainty. We aim to rectify this situation by using probabilistic planning. The first step is to create an algorithm that can generate many realistic contours with variations that match actual inter-observer variation. Such an algorithm is developed here.

**Materials/Methods**
Prostate delineations from five observers on 18 MR datasets, belonging to the golden atlas [1], were used. For each patient, a median shape from these five delineations were created following the methodology described in [2]. We assume hereafter that the median shape represents the ground truth. Signed distance maps (negative values inside and positive outside) were created for each observer and the median shape ($d_{median}$) in an 1x1x1 mm$^3$ grid. The observers' distance maps were used to calculate an inter-observer variability map, *IOV*, by calculating the standard deviation of the five distance maps per voxel.

Contour generation: We adopted a level set formulation of the median shape, with the level set function defined as $d_{median}$. A noise map, $\eta$, was generated by sampling values from a normal distribution $\mathcal{N}(\mu = 0, \sigma^2 = 1)$ on the same grid as $d_{median}$. The noise map was then weighted by multiplying it with the IOV map, $\eta' = \eta \cdot IOV$. To account for the geometrical dependency of IOV uncertainties, the weighted noise was smoothed by applying a gaussian blur with standard deviation $\sigma$, $\eta'_\sigma = \eta' * G(\sigma)$., i.e., representing a point spread function (PSF). The new contour was composed by all points $P = \{x|f(x) = 0\}$, where $f(x) = d_{median} + \eta'_\sigma$. Figure 1 shows the effect of using different $\sigma$ values.

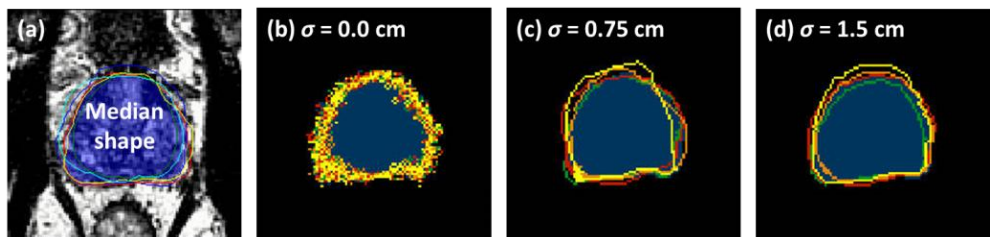

**Figure 1:** *(a) Input data for patient 1_01;(b-d) contours generated with our new algorithm using different values for σ, i.e. the PSF. The median shape is represented as a blue shadow in all panels.*

Since inter-observer variability is often assumed homogeneous, we also explored using a constant IOV map. We used the root mean square of all standard deviations on the median shape surface per patient.

Contour similarity evaluation: we aimed at generating contours that are geometrically similar to the manually drawn contours, and that produce similar dosimetric variations assuming the plan was based on the median shape. We generated 100 contours per IOV and constant map, varying $\sigma$ between 0 and 1.5 cm, resulting in 1400 contours per patient. To calculate the geometrical similarity of the contours, we measured the maximum distance to agreement (maxDTA) between the contours and the median. To assess the dosimetric variations, we constructed an ideal dose distribution following the median shape, with a penumbra of 0.32 cm [3]. We then collected the minimum dose in each contour, minDose. We compared the distributions of maxDTA and minDose between the manually drawn contours and generated contours using the two-sample Kolmogorov–Smirnov test. Note that the Kolmogorov–Smirnov test is devised to be sensitive against all possible types of differences between two distribution functions.

**Results**

Contours created using $\sigma = 0\ cm$ were unrealistic, i.e., it is essential to take the PSF into account. Figure 2 shows the distributions of maxDTA and minDose for all contours. Using $\sigma = 0.75\ cm$ and the (non-homogenous) IOV map produced maxDTA and minDose distributions that were not significantly different from the distributions of manually drawn contours (p=0.085 and p=0.607). All other distributions were significantly different (p<0.05).

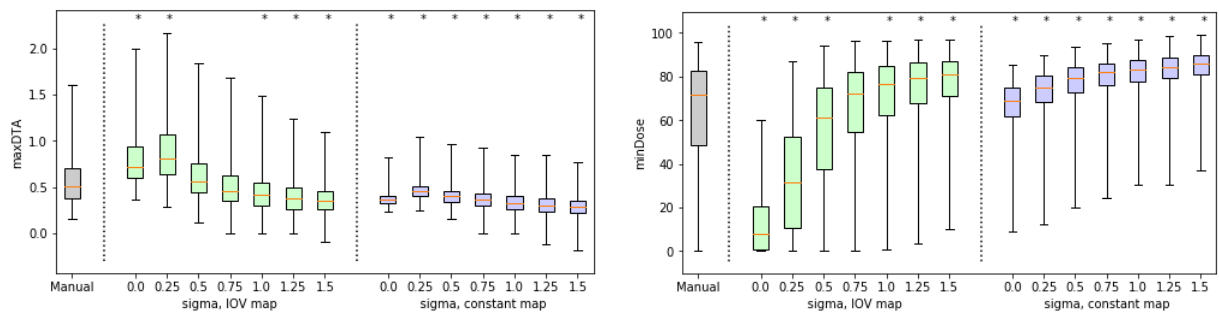

**Figure 2:** *Box-plot showing the distributions of maxDTA and minDose for all contours: manually drawn contours in grey, generated contours using the IOV maps in green and using the constant map in light blue. Distributions significantly different to the manually drawn contours are marked with a \* (p<0.05). Note red line is median, the box shows inter-quartile range and whisker show the range of the data.*

**Discussion/Conclusions**

We successfully developed an algorithm that generates contours that match geometrical and dosimetric characteristics of manually delineated contours. In our data, using a PSF with $\sigma = 0.75\ cm$ produced maxDTA and minDose distributions that were not significantly different from the manually delineated structures. Accounting for the PSF is essential to correctly simulate inter-observer variation. The first step to incorporate contour variations directly into treatment optimisation has been taken.

**References**

[1] T. Nyholm, et al., *MR and CT data with multiobserver delineations of organs in the pelvic area—Part of the Gold Atlas project. Med. Phys.*, 45: 1295-1300. doi:10.1002/mp.12748, 2018
[2] K. E. Deurloo, et al., *Quantification of shape variation of prostate and seminal vesicles during external beam radiotherapy*, IJROBP, 61(1): 228-238, 2005.
[3] M van Herk, et al., *The probability of correct target dosage: dose-population histograms for deriving treatment margins in radiotherapy.* IJROBP, 47(4):1121-35, 2000.

**Acknowledgements**: This work is supported by the NIHR Manchester Biomedical Research Centre.

[†]Corresponding author: eliana.vasquezosorio@manchester.ac.uk